\documentclass[11pt,a4paper]{article}
\usepackage{amssymb}
\usepackage{epsf}
\newlength{\HFPP}       \HFPP5.4mm
\makeatletter
\@addtoreset{equation}{section}
\makeatother

\def\preprint#1#2{\noindent\hbox{#1}\hfill\hbox{#2}\vskip 10pt}
\def\k_tilde{\sqrt{\frac{\kappa}{1-\kappa}}}
\def\kt2{\frac{\tilde{\kappa}}{2}}

\begin{document}
\begin{titlepage}
\def\thefootnote{\fnsymbol{footnote}}

\preprint{ITP-UH-5/97}{February 1997}
\vfill

\begin{center}
  {\Large\sc Spectrum of boundary states in the open Hubbard chain}
\vfill

{\sc Gerald Bed\"urftig}\footnote{e-mail: bed@itp.uni-hannover.de}
	and
{\sc Holger Frahm}\footnote{e-mail: frahm@itp.uni-hannover.de}
\vspace{1.0em}

{\sl
  Institut f\"ur Theoretische Physik, Universit\"at Hannover\\
  D-30167~Hannover, Germany}
\end{center}
\vfill

\begin{quote}
  We use the Bethe Ansatz solution for the one dimensional Hubbard model
  with open boundary conditions and applied boundary fields to study the
  spectrum of bound states at the boundary.  Depending on the strength of
  the boundary potentials one finds that the true ground state contains a
  single charge or, for boundary potentials comparable to the Hubbard
  interaction, a pair of electrons in a bound state.  If these are left
  unoccupied one finds holon and spinon bound states.  We compute the
  finite size corrections to the low lying energies in this system and use
  the predictions of boundary conformal field theory to study the exponents
  related to the orthogonality catastrophe.
\end{quote}

{PACS-Nos.: 
05.70.Jk,	
71.10.Fd,	
71.10.Pm,	
73.20.-r	
}
\vspace*{\fill}
\setcounter{footnote}{0}
\end{titlepage}

%
%
\section{Introduction}
The recent advances in the understanding of boundary effects in low
dimensional quantum systems due to the predictions of boundary conformal
field theory \cite{card:89,aflu:94,affl:94} and the formulation of Bethe
Ansatz soluble models on open lattices with potentials applied on the
boundary sites \cite{hschulz:85,alca:87,skly:88,menr:90} have opened new
possibilities to study the effects of correlations and quantum fluctuations
on long standing problems such as the orthogonality catastrophe
\cite{and:67,qfyu:96} and edge singularities in optical absorption
experiments \cite{nodo:69,scho:69,affl:96}.

The effect of electronic correlations on the bulk critical behaviour of
$(1+1)$-dimensional quantum systems has been studied successfully in the
Tomonaga-Luttinger model which then can be handled using field theoretical
methods \cite{Tomonaga,Luttinger,lupe:75}.  Studies of integrable lattice
models have added insights to this problem since e.g.\ the dependence of
critical exponents on microscopic parameters and their behaviour due to
lattice effects (back scattering, Mott transition) can be computed exactly
\cite{frko:90,frko:91,kaya:91}.  Similarly, one expects additional
information from studies of lattice models for interacting electrons with
open boundaries \cite{hschulz:85,assu:96,essl:96}.  Besides giving a deeper
understanding of previous predictions these lattice models have features
not easily included into the continuum description: local chemical
potentials in the former lead to a sequence of bound states (see e.g.\
\cite{kask:96}) which are expected to influence the critical properties of
the boundary.

In this paper we consider the Hubbard model on a chain of $L$ sites subject
to an additional chemical potential $p$ at the first site.  The Hamiltonian
is given by
\begin{equation}
  {\cal H}=-\sum_{\sigma,j=1}^{L-1}
       \left( c_{j,\sigma}^\dagger c_{j+1,\sigma}+h.c.\right)
       +4u \sum_{j=1}^{L}n_{j \uparrow}n_{j \downarrow}+\mu \hat{N}
       -{h \over 2} (\hat{N}-2\hat{N}_\downarrow)
       -p (\hat{N}_{1,\uparrow}+\hat{N}_{1,\downarrow}).
\label{ham:hubb}
\end{equation}
For $p=0$ this model has been solved by Schulz using the coordinate Bethe
Ansatz \cite{hschulz:85}.  Only recently, this solution has been extended
to nonvanishing $p$ \cite{assu:96} and the integrability of the model has
been established in the framework of the quantum inverse scattering method
\cite{zhou:96}.  The Bethe Ansatz equations (BAE) determining the spectrum
of (\ref{ham:hubb}) in the $N_{e}$-particle sector with magnetization
$M={1\over2} N_{e}-N_{\downarrow}$ read \cite{assu:96}
\begin{eqnarray}
e^{ik_j 2(L+1)}s_p(k_j)&=&\prod_{\beta=1}^{N_\downarrow}
e_{2u}(\eta_j-\lambda_\beta)
e_{2u}(\eta_j+\lambda_\beta)
\qquad j=1,\ldots,N_e \nonumber \\
\prod_{j=1}^{N_e} 
e_{2u}(\lambda_\alpha-\eta_j)
e_{2u}(\lambda_\alpha+\eta_j) &=&
\prod_{\beta=1 \atop \beta \not= \alpha}^{N_\downarrow}
e_{4u}(\lambda_\alpha-\lambda_\beta)
e_{4u}(\lambda_\alpha+\lambda_\beta)
\qquad \alpha=1,\ldots,{N_\downarrow}. \nonumber \\
\label{eq:bae}
\end{eqnarray}
with $e_n(x)=\frac{x+i{n \over 2}}{x-i{n \over 2}}$, $s_p(k_j)=
\left(\frac{1-pe^{-ik_j}}{1-pe^{ik_j}}\right)$ and $\eta_j=\sin{}k_j$.  The
energy of the corresponding eigenstate of (\ref{ham:hubb}) is
\begin{equation}
   E=\sum_{j=1}^{N_e}\left(\mu-{h \over 2}-2 \cos{}k_j\right)
	+h N_\downarrow\ .
\label{energy}
\end{equation}
Using the global spin- and $\eta$--pairing SU(2) symmetry of the Hubbard
model the Bethe states extended by those obtained by application of the
corresponding raising operators have been shown to form a complete basis of
the Hilbert space of the system \cite{eks:92}.  A non-zero boundary
potential destroys the $\eta$--symmetry of the model and the question of
completeness should be considered again.  Numerical solutions of
(\ref{eq:bae}) for small $p$ show that there exist complex combinations of
two $k$ and one $\lambda$ which coincide with one $\eta$--pair in the limit
of $p \to 0$.  In the following we only consider the ground state and the
low lying excitations of the system, so we can neglect these kind of
complex solutions as they belong to the highly excited states of the system
\cite{yang:89}.  However, for sufficiently strong attractive boundary
potentials $p>1$ we find that there exist other complex solutions which
turn out to correspond to bound states in these potentials (note that these
states do not appear in the case $p<1$ studied in \cite{assu:96}).  These
solutions need to be considered to obtain the true ground state of the
system.  We find that in spite of the presence of several complex
parameters in the ground state configuration the low--energy spectrum of
the many particle system can still be described in the Tomonaga--Luttinger
picture equivalent to two $c=1$ Conformal Field Theories.  The case $p>1$
will be studied in detail in the next section.
\section{Boundary bound states}
From a physical point of view it is clear that the ground state of the
model contains a bound state at the first site for sufficiently large $p$.
Numerical solutions of the BAE show that this is indeed the fact for $p>1$
where a complex quasi momentum $k$ is present in the ground state
configuration.  A similar situation has been found in the XXZ Heisenberg
chain with a boundary magnetic field \cite{kask:96,skosa:95} and in a
continuum model related to the Kondo problem \cite{wavo:96}.

Increasing the boundary potential further we find that additional complex
parameters are added to the gound state solution of (\ref{eq:bae}).  In the
thermodynamic limit $(L \to \infty)$ we have to distinguish three different
regions where the BAE describing the ground state are modified due to the
presence of these complex roots:\footnote{%
In principle one is free to leave the bound states empty.  This gives rise
to another continuum of states.  These states become important if one
considers e.g.\ multiple Fermi edge singularities in the presence of bound
states \cite{affl:96,esfr:pp}.}
\begin{itemize}
\item[{\bf I}:]{$1<p<p_1=u+\sqrt{1+u^2}$}
\begin{eqnarray}
  &&e^{ik_j 2(L+1)}s_p(k_j)=\prod_{\beta=1}^{N_\downarrow}
    e_{2u}(\eta_j-\lambda_\beta)
    e_{2u}(\eta_j+\lambda_\beta)\ ,
    \qquad j=1,\ldots,N_e-1 
\nonumber \\
  &&e_{2u-2t}(\lambda_\alpha)
    e_{2u+2t}(\lambda_\alpha)
    \prod_{j=1}^{N_e-1} 
    e_{2u}(\lambda_\alpha-\eta_j)
    e_{2u}(\lambda_\alpha+\eta_j)
   =   \prod_{\beta=1 \atop \beta \not= \alpha}^{N_\downarrow} 
    e_{4u}(\lambda_\alpha-\lambda_\beta)
    e_{4u}(\lambda_\alpha+\lambda_\beta)\ , 
\nonumber \\
   &&\qquad \alpha=1,\ldots,{N_\downarrow}
\label{eq:c1}
\end{eqnarray}
with the complex solution $k_{N_e} = i \ln(p)$ (with exponential accuracy
in the limit $L \to \infty$) and $t=-i\sin{}k_{N_e}={1\over 2}(p-{1\over
p})<u$.  The contribution of this bound state to the energy (\ref{energy})
is given by $E_1=-p-{1 \over p}+\mu-{h \over 2}$. This complex solution
corresponds to a charge bound to the first site, as the quasimomenta $k$
parametrize the charge part of the states.

\item[{\bf II}:]{$p_1<p<p_2=2u+\sqrt{1+4u^2}$}\\
Larger values of the boundary potential lead to an additional complex
solution in the spin part: $\lambda_{N_\downarrow}=i(t-u)$ ($t>u$ in this
region) and the following modified BAE:
\begin{eqnarray}
  &&e^{ik_j 2(L+1)}s_p(k_j)=e_{4u-2t}(\eta_j)e_{2t}(\eta_j)
	\prod_{\beta=1}^{{N_\downarrow}-1}
	e_{2u}(\eta_j-\lambda_\beta)
	e_{2u}(\eta_j+\lambda_\beta)\ ,
	\qquad j=1,\ldots,N_e-1 
\nonumber \\
  &&\prod_{j=1}^{N_e-1} e_{2u}(\lambda_\alpha-\eta_j)
	e_{2u}(\lambda_\alpha+\eta_j)
	=e_{2t-2u}(\lambda_\alpha)e_{6u-2t}(\lambda_\alpha)
	\prod_{\beta=1 \atop \beta \not= \alpha}^{{N_\downarrow}-1}
	e_{4u}(\lambda_\alpha-\lambda_\beta)
	e_{4u}(\lambda_\alpha+\lambda_\beta)\ ,
\nonumber\\
	&&\qquad \alpha=1,\ldots,{N_\downarrow}-1
\label{eq:c2}
\end{eqnarray}
Again, this state can be interpreted as that of a charge bound to the
surface.  The physical excitations in the spin sector --- so called {\em
spinons} --- correspond to {\em holes} in the distribution of spin
rapidities $\lambda$ which are still real.

\item[{\bf III}:]{$p>p2$}\\
For boundary potentials larger than the Hubbard interaction $p\gtrapprox4u$
a pair of electrons forming a singlet is bound to the surface, parametrized
by $\lambda_{N_\downarrow}=\sin{}k_{N_e}\ -iu =\sin{}k_{N_e-1}\
+iu=i(t-u)$.  The resulting BAE are
\begin{eqnarray}
e^{ik_j 2(L+1)}s_p(k_j)e_{2t-4u}(\eta_j)
&=&
e_{2t}(\eta_j)
\prod_{\beta=1}^{{N_\downarrow}-1}
e_{2u}(\eta_j-\lambda_\beta)
e_{2u}(\eta_j+\lambda_\beta)
\qquad j=1,\ldots,N_e-2 \nonumber \\
\prod_{j=1}^{N_e-2} 
e_{2u}(\lambda_\alpha-\eta_j)
e_{2u}(\lambda_\alpha+\eta_j) &=&
\prod_{\beta=1 \atop \beta \not= \alpha}^{{N_\downarrow}-1}
e_{4u}(\lambda_\alpha-\lambda_\beta)
e_{4u}(\lambda_\alpha+\lambda_\beta)
\qquad \alpha=1,\ldots,{N_\downarrow}-1 \nonumber \\
\label{eq:c3}
\end{eqnarray}
The energy of the second complex solution $k_{N_e-1}$ is given by 
$E_2=-2 \sqrt{1+(t-2u)^2}+\mu-{h \over 2}$.
\end{itemize}
Note that region I is already realized in the ferromagnetic case with
spin-$\uparrow$ electrons only. As $t=u$ ($p=p_1$) the index of the first
factor in the $\lambda$--equation of (\ref{eq:c1}) changes the sign,
allowing for the complex $\lambda$--solution.  A similar change occurs in
(\ref{eq:c1}) for $t=2u$ ($p=p2$) leading to the second complex
$k$--solution.  No such point exists in (\ref{eq:c3}), hence no further
complex solutions are expected in the ground state --- in perfect agreement
with the physical intuition.

Recently the BAE for the model with a boundary magnetic field
$(-p_1(n_{1\uparrow}-n_{1\downarrow}))$ applied at the first site have been
constructed \cite{shwa:97,deyu:pp}.  This field induces an additional phase
factor $-e_{2u-2t}(\lambda_\alpha)$ in the second eq.\ of (\ref{eq:bae})
which cancels the first factor in (\ref{eq:c1}) (up to a sign).  As a
consequence, we do not expect another complex solution to exist in the
ground state besides the first one for this case.

Using standard procedures the BAE for the ground state and low lying
excitations can be rewritten as linear integral equations for the densities
$\rho_c(k)$ and $\rho_s(\lambda)$ of real (positive) quasi momenta $k_j$
and spin rapidities $\lambda_\alpha$, respectively.  Identification of
positive and negative $k$ and $\lambda$ allows to symmetrize the resulting
equations with the usual result
\begin{equation}
   \left( \begin{array}{c} \rho_c \\ 
			   \rho_s \end{array} \right) =
   \left( \begin{array}{c} {1 \over \pi}+{1 \over L}\hat{\rho}_c^0\\ 
			   {1 \over L}\hat{\rho}_s^0 \end{array} \right)+
   \left( \begin{array}{cc} 0 & \cos{}k\ a_{2u}(\eta-\lambda') \\ 
	                    a_{2u}(\lambda-\eta') & 
		-a_{4u}(\lambda-\lambda') \end{array} \right)*
   \left( \begin{array}{c} \rho_c \\ 
			   \rho_s \end{array} \right)\ .
 \label{eq:dnorm}
\end{equation}
Here we have introduced $a_y(x)={1 \over 2\pi}\frac{y}{y^2/4+x^2}$ and
$f*g$ denotes the convolution $\int_{-A}^{A}dy f(x-y)g(y)$ with the
boundaries $k_0$ and $\lambda_0$ in the charge and spin sector,
respectively.  The latter are fixed by the conditions
\begin{equation}
   \int_{-k_0}^{k_0}dk \rho_c =
	\frac{2\left[N_e-\theta(p-1)-\theta(p-p_2)\right]+1}{L}\quad,
 \quad
   \int_{-\lambda_0}^{\lambda_0}d\lambda\rho_s =
	\frac{2\left[N_\downarrow-\theta(p-p_1)\right]+1}{L},
\label{eq:fix}
\end{equation}
where $\theta(x)$ is the Heaviside step function.  The driving terms of the
$1/L$-corrections in the different regions are given by
\begin{equation}
  \hat{\rho}_c^0(k)= 
	{1 \over \pi}-\cos{}k\ a_{2u}(\eta)
		+\frac{\cos{}k\ p-p^2}{\pi(p^2+1-2p \cos{}k)}
	+\theta(p-p1)\cos{}k\left[a_{2t}(\eta)+a_{4u-2t}(\eta)\right] 
\end{equation}
for the charge--sector\footnote{%
Note that the index of $a_{4u-2t}$ changes sign at $p=p_2$.}
and
\begin{equation}
  \hat{\rho}_s^0(\lambda)= 
	a_{4u}(\lambda)+
 \left\{
	\begin{array}{ll}
	0 & p<1 \\
	a_{2u-2t}(\lambda)+a_{2u+2t}(\lambda) & I\\
	-a_{2t-2u}(\lambda)-a_{6u-2t}(\lambda)& II\\
	0& III  
	\end{array}\right.
\end{equation}
for the spin--sector. In terms of the dressed energies $\varepsilon_c$ and
$\varepsilon_s$ which satisfy the same integral equations as in the Hubbard
model with periodic boundary conditions:
\begin{equation}
   \left( \begin{array}{c} \varepsilon_c \\ 
			   \varepsilon_s \end{array} \right) =
   \left( \begin{array}{c} \mu-{h \over 2}-2 \cos{}k\\ 
			   h \end{array}\right)+
   \left( \begin{array}{cc} 0 & a_{2u}(\eta-\lambda') \\ 
	                    a_{2u}(\lambda-\eta') \cos{}k'& 
		-a_{4u}(\lambda-\lambda') \end{array} \right)*
   \left( \begin{array}{c} \varepsilon_c \\ 
			   \varepsilon_s \end{array} \right)
 \label{eq:denorm}
\end{equation}
the energy of the state can be expressed as:
\begin{eqnarray}
{E \over L} &=&  e_\infty + {1 \over L} f_\infty +o\left({1 \over L}\right) 
 = {1 \over 2}\int\limits_{-k_0}^{k_0}dk\; 
	{\varepsilon_c(k)\over \pi}+ 
{1 \over 2L}\left[\int\limits_{-k_0}^{k_0}dk\; 
	\varepsilon_c(k)\hat{\rho}_c^0(k)+
\int\limits_{-\lambda_0}^{\lambda_0}d\lambda\; 
	\varepsilon_s(\lambda)\hat{\rho}_s^0(\lambda)\right]
\nonumber \\ &+&
{1 \over 2L}\left[
-(\mu+{h \over 2}-2)
+2\theta(p-1)E_1+2h\theta(p-p_1)
+2\theta(p-p_2)E_2 \right]+o\left({1 \over L}\right)\ .
	\nonumber \\
\label{eq:enb1l}
\end{eqnarray}

\section{Ground state expectation value of $N_1$}
\begin{figure}[ht]
(a)
\epsfxsize=0.46\textwidth
\epsfbox{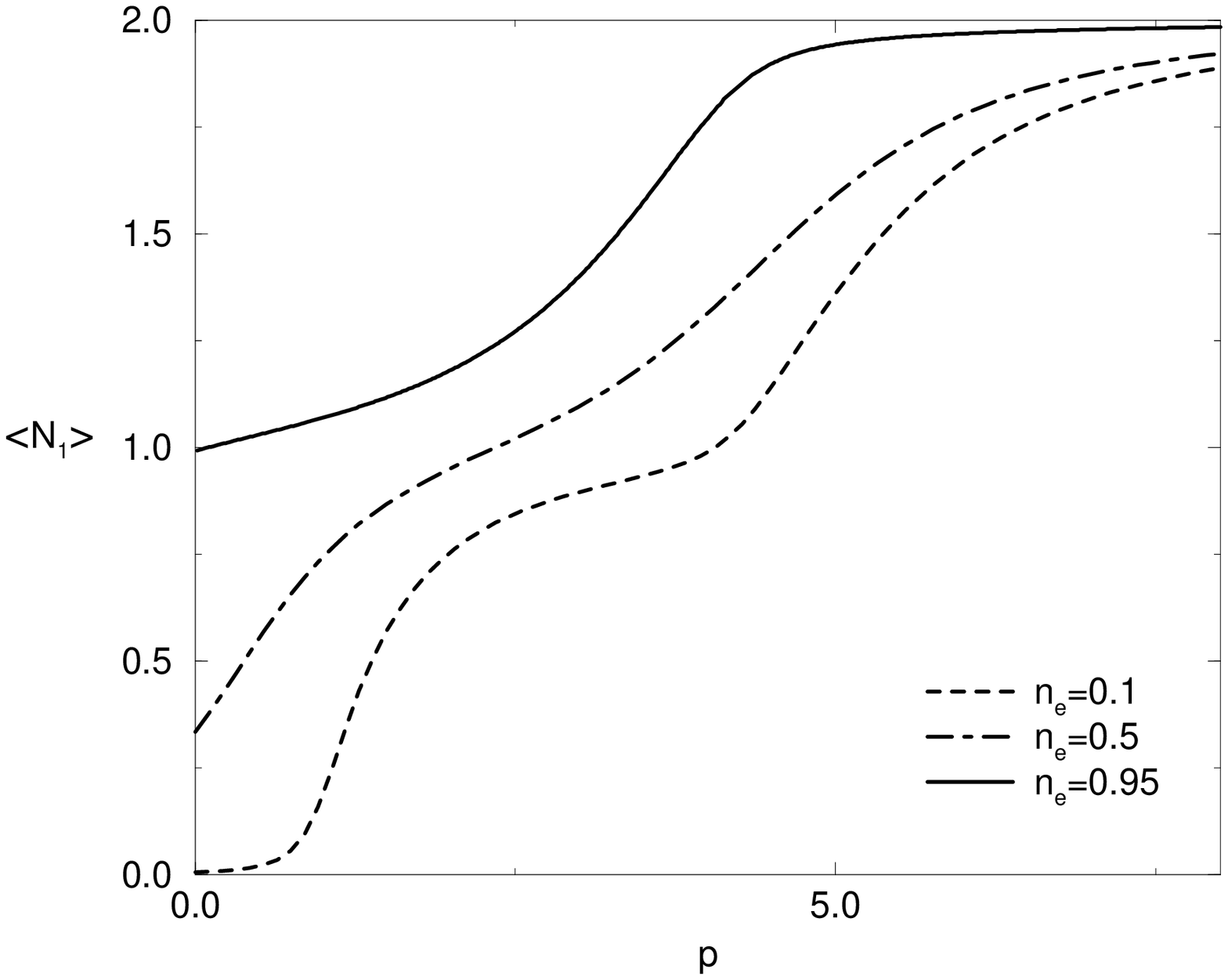}
\hfill
(b)
\epsfxsize=0.46\textwidth
\epsfbox{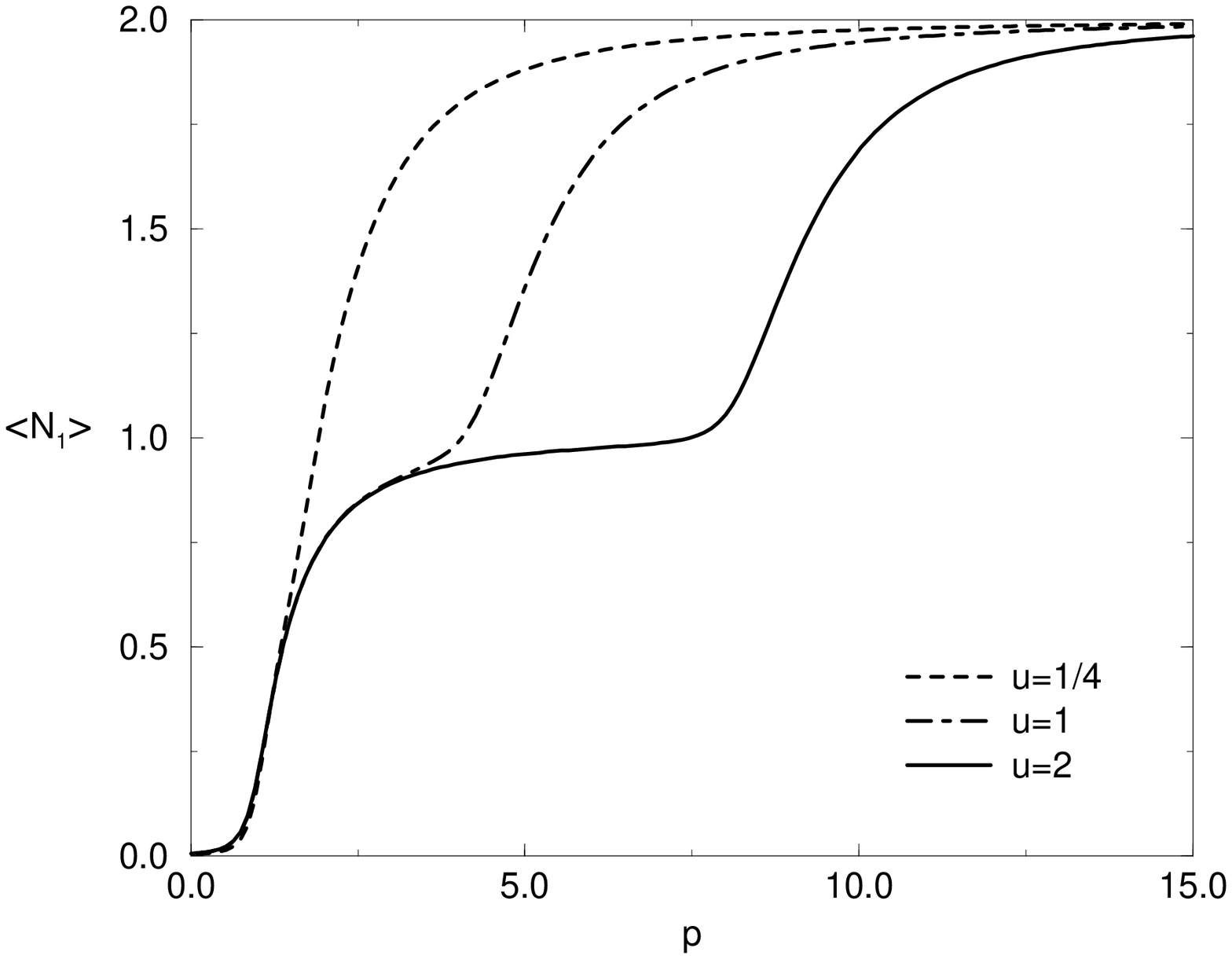}\\
(c)
\epsfxsize=0.46\textwidth
\epsfbox{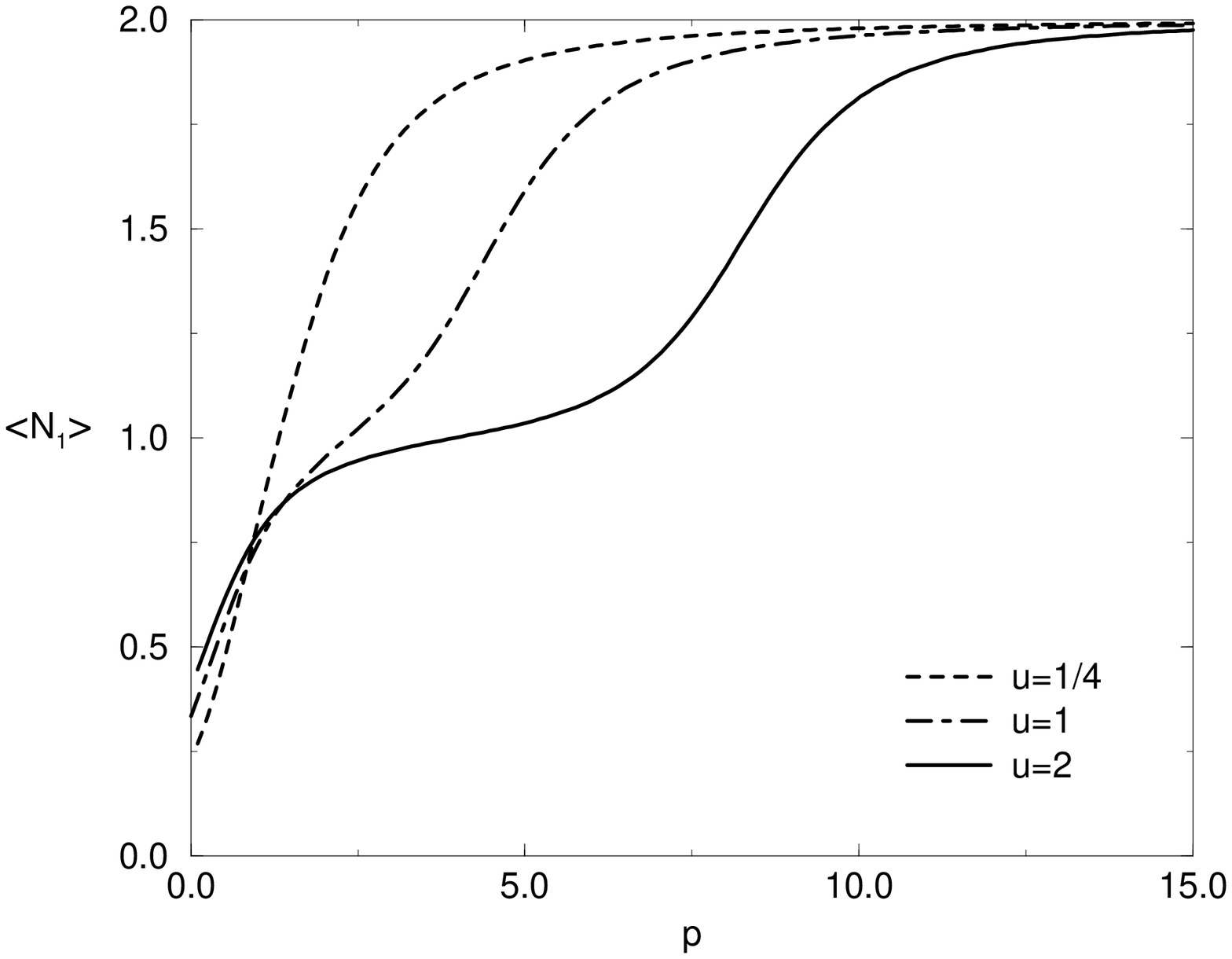}
\hfill
(d)
\epsfxsize=0.46\textwidth
\epsfbox{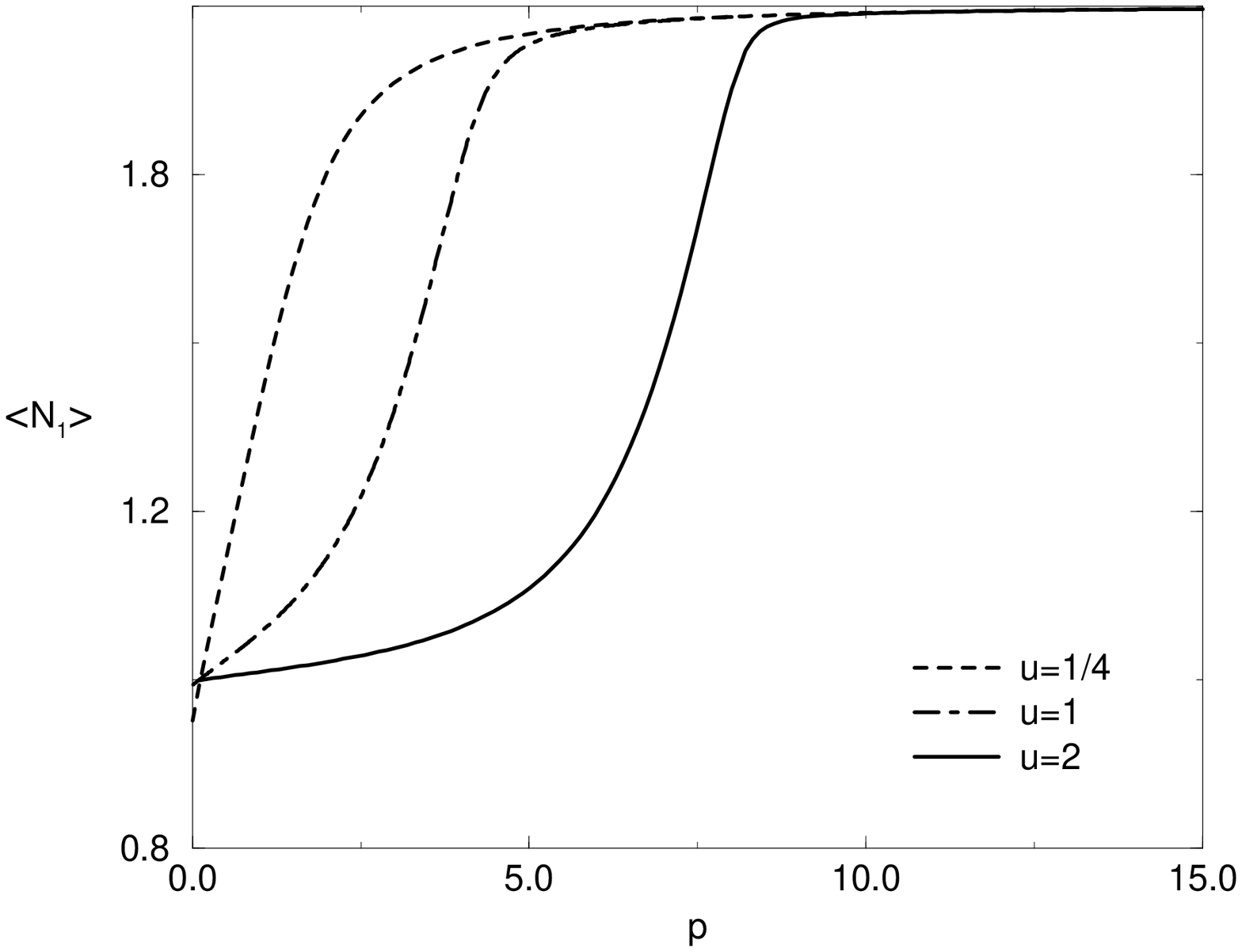}
\caption{ Ground state expectation value of $N_1$ as a function of the
boundary potential $p$ for
$u=1$ and several electron densities $n_e$ (a);
fixed density $n_e=0.1$ (b), $n_e=0.5$ (c), $n_e=0.95$ (d) and several
values of $u$.
\label{fig:U4}}
\end{figure}

The ground state expectation values for the occupation of the boundary site
$\langle N_1 \rangle$ can be calculated from the identity $\langle N_1
\rangle =-{\partial E}/{\partial p}$.  With (\ref{eq:enb1l}) we obtain
\begin{equation}
  \langle N_1 \rangle=-{1 \over 2}\left[\int\limits_{-k_0}^{k_0}dk\; 
  \varepsilon_c(k)\frac{\partial \hat{\rho}_c^0(k)}{\partial p}+
  \int\limits_{-\lambda_0}^{\lambda_0}d\lambda\; 
  \varepsilon_s(\lambda)\frac{\partial\hat{\rho}_s^0(\lambda)}{\partial p}
  +2\theta(p-1)\frac{\partial E_1}{\partial p}+2\theta(p-p_2)
  \frac{\partial E_2}{\partial p}\right]
\end{equation}
In absence of a bulk magnetic field $h$ the ground state of the Hubbard
model is known to be a singlet (for even particle number) corresponding to
$\lambda_0=\infty$.  In this case the system of integral equations
(\ref{eq:denorm}) can be reduced to a scalar one
\begin{equation}
  \varepsilon_c(k)=\mu-2 \cos{}k +\int_{-k_0}^{k_0}dk'\;
  G_{2u}^{2u}(\eta-\eta')\cos{}k'\ \varepsilon_c(k')
\end{equation}
with ($y>0$, $y+z>0$)
\begin{equation}
  G_{y}^{z}(\lambda)={1 \over 2\pi y}\hbox{Re}\left\{\Psi\left({3 \over 4}
  +{z \over 4y}+i{\lambda \over 2 y}\right)-\Psi\left({1 \over 4}+
  {z \over 4y}+i{\lambda \over 2 y}\right)
  \right\}
  ,\quad G_y^{z}(\omega)=\frac{e^{-{z \over 2}|\omega|}}
  {2\cosh\left({y \over 2}\omega\right)}
\end{equation}
($\Psi$ is the digamma function).  We obtain
\begin{eqnarray}
   \langle N_1 \rangle&=&
   -\theta(p-1)\frac{\partial E_1}{\partial p}-\theta(p-p_2)
   \frac{\partial E_2}{\partial p}
\nonumber\\
   &&-{1 \over 2}\int_{-k_0}^{k_0}dk\;
   \varepsilon_c(k)
	\left\{
	\begin{array}{ll}
	\gamma_p(k) & p<1 \\
	\gamma_p(k)+\frac{\partial}{\partial p}
	\left(G_{2u}^{2u-2t}(\eta)+G_{2u}^{2u+2t}(\eta)\right)\cos{}k& I,II\\
	\gamma_p(k)+\frac{\partial}{\partial p}
	\left(a_{2t}(\eta)-a_{2t-4u}(\eta)\right)\cos{}k & III
	\end{array}\right.
\end{eqnarray}
with $\gamma_p(k)=\frac{\cos{}k\ p^2+\cos{}k -2p}{\pi(p^2+1-2\cos{}k\
p)^2}$.  In the limit of $p \to \infty$ only the the first two parts
survive and we get the expected result $\langle N_1 \rangle=2$.  Some
numerical results are shown in Fig.~\ref{fig:U4}.

\section{Finite size corrections}
Following \cite{woyn:89} we can calculate the finite size spectrum of the
model, reproducing the result of \cite{assu:96}:
\begin{eqnarray}
  E = L e_\infty + f_\infty &+&
        {\pi v_c \over{L}} \left\{ -{1\over24} + {1\over{2\det^2(Z)}} 
                \left[\left(\Delta N_c^0 -\theta^c_p\right)Z_{ss}
		- \left(\Delta N_s^0 -\theta^s_p\right)Z_{cs}\right]^2
		+N_c^+\right\}
\nonumber \\
        &+& {\pi v_s \over{L}}\left\{-{1\over24} +{1\over{2\det^2(Z)}} 
		\left[\left(\Delta N_s^0 -\theta^s_p\right)Z_{cc}
		- \left(\Delta N_c^0 -\theta^c_p\right)Z_{sc}\right]^2
		+N_s^+\right\}.
\label{fsopen:h}
\end{eqnarray}
Here $N_{c,s}^+$ are non negative integers counting the number of particle
hole excitations at the Fermi points, the Fermi velocities are given by
$v_{c}={\varepsilon'_{c}(k_0)\over\pi\rho_{c}(k_0)}$ and
$v_{s}={\varepsilon'_{s}(\lambda_0)\over\pi\rho_{s}(\lambda_0)}$.  $Z$ is
the dressed charge matrix
\begin{equation}
Z= \left( \begin{array}{cc} Z_{cc} & Z_{cs} \\ 
  Z_{sc} & Z_{ss} \end{array} \right)=
  \left( \begin{array}{cc} \xi_{cc}(k_0) & \xi_{sc} (k_0) \\ 
  \xi_{cs} (\lambda_0) & \xi_{ss}(\lambda_0) \end{array} \right)^\top
\label{eq:z}
\end{equation}
given in terms of the integral equation
\begin{equation}
\left( \begin{array}{cc} \xi_{cc}(k) & \xi_{sc}(k) \\ 
  \xi_{cs}(\lambda) & \xi_{ss}(\lambda) \end{array} \right)=
\left( \begin{array}{cc} 1 & 0 \\ 0 & 1 \end{array} \right)+
 \left( \begin{array}{cc} 0 & a_{2u}(\eta-\lambda') \\ 
a_{2u}(\lambda-\eta')\cos{}k' & -a_{4u}(\lambda-\lambda') \end{array} \right)*
 \left( \begin{array}{cc} \xi_{cc}(k') & \xi_{sc}(k') \\ 
  \xi_{cs}(\lambda') & \xi_{ss}(\lambda') \end{array} \right)\ .
\end{equation}
The $\Delta N^0_{c,s}$ are given by $\Delta N^0_c=N_e-L n_e$ and $\Delta
N^0_s=N_\downarrow-{L}n_\downarrow$, where $n_e$ and $n_\downarrow$ denote
the total density of electrons and spin-$\downarrow$ electrons of the
reference state which we define through
\begin{equation}
   n_e={1\over2}\int_{-k_0}^{k_0}dk \rho_c^{(0)}(k)\ ,
   \qquad
   n_\downarrow={1\over2}\int_{-\lambda_0}^{\lambda_0}
	d\lambda\rho_s^{(0)}(\lambda)\ .
\label{eq:fix2}
\end{equation}
Here $\rho_{cs}^{(0)}$ should be computed from (\ref{eq:dnorm}) {\em
without} the $1/L$ terms, i.e.\ $\hat{\rho}_{cs}^0\equiv 0$ (note that this
choice differs from that used in \cite{assu:96}).  This choice implies that
for a given boundary condition $\Delta N^0_{c,s}=\theta_p^{c,s}$ are
nonzero in the corresponding ground state.  The shifts $\theta_p^{c,s}$ are
due to the ${1 \over L}$--terms in (\ref{eq:dnorm}):
\begin{eqnarray}
\theta^c_p&=&{1 \over 2}\left(\int_{-k_0}^{k_0}dk \hat{\rho}_c-1+2\theta(p-1) 
+2\theta(p-p2)\right) \nonumber \\ 
\theta^s_p&=&{1 \over 2}\left(\int_{-\lambda_0}^{\lambda_0}d\lambda \hat{\rho}_s
-1+2\theta(p-p1)\right)
\label{eq:tpc}
\end{eqnarray}
with $\hat{\rho}_c$ and $\hat{\rho}_s$ denoting the solution of
(\ref{eq:dnorm}) without the ${1 \over \pi}$ driving term.  Hence the
finite size spectrum (\ref{fsopen:h}) determinig the bulk correlation
functions \cite{frko:90} can be written in a manifestly particle-hole
symmetric form by introducing $\Delta \widetilde{N}_{c,s}^0=\Delta
N_{c,s}^0+\theta^{c,s}_p$, where $\Delta \widetilde{N}_{c,s}^0$ denotes the
change in charge and spin as compared to the ground state (see also
\cite{fuka:96,wavp:96}):
\begin{eqnarray}
  E = L e_\infty + f_\infty &+&
        {\pi v_c \over{L}} \left\{ -{1\over24} + {1\over{2\det^2(Z)}} 
                \left[\Delta \widetilde{N}_c^0Z_{ss}
		- \Delta \widetilde{N}_s^0 Z_{cs}\right]^2
	+N_c^+\right\}
\nonumber \\
        &+& {\pi v_s \over{L}}\left\{-{1\over24} +{1\over{2\det^2(Z)}} 
		\left[\Delta \widetilde{N}_s^0 Z_{cc}
		- \Delta \widetilde{N}_c^0 Z_{sc}\right]^2+N_s^+\right\}.
\label{fsopentil:h}
\end{eqnarray}

These expressions simplify in certain limits (see also the corresponding
discussion for the periodic model in \cite{frko:90,frko:91}):\\
\underline{\sc Zero magnetic field ($\lambda_0=\infty$)}: 
The spin part of the equations can be eliminated by Fourier transformation
with the result that the matrix $Z$ depends on the scalar dressed charge
$\xi=\xi(k_0)$ only \cite{woyn:89}:
\begin{equation}
Z= \left( \begin{array}{cc} Z_{cc} & Z_{cs} \\ 
  Z_{sc} & Z_{ss} \end{array} \right)=
  \left( \begin{array}{cc} \xi & 0 \\ 
  {1\over 2}\xi & {\sqrt{2} \over 2} \end{array} \right) \label{eq:zhn}
\end{equation}
which is defined as the solution of 
\begin{equation}
\xi(k)=1+\int_{-k_0}^{k_0} dk' \cos{}k'\ G_{2u}^{2u}(\eta-\eta') \xi(k'). 
\label{eq:xi}
\end{equation}
Furthermore, one finds the relation $\theta^s_p={1\over2}\theta^c_p$, which
allows to rewrite the finite size spectrum (\ref{fsopen:h}) as
\begin{eqnarray}
  E = L e_\infty + f_\infty 
     &+& {\pi v_c \over{L}} \left\{ -{1\over24} + {1\over{2\xi^2}} 
                \left(\Delta N_c^0 -\theta^c_p\right)^2 +N_c^+\right\}
\nonumber\\
     &+& {\pi v_s \over{L}}\left\{-{1\over24} +
                \left(\Delta N_s^0-{1\over2}\Delta N_c^0\right)^2
		+ N_s^+ \right\}\ .
\label{fsopen:h0}
\end{eqnarray}
The function $\hat{\rho}_c$ in (\ref{eq:tpc}) satisfies the integral
equation $\hat{\rho}_c(k)= \tilde{\rho}_c(k)+\cos{}k\
\int_{-k_0}^{k_0}dk'\; G_{2u}^{2u}(\eta-\eta')\hat{\rho}_c(k')$ with
driving term
\begin{equation}
  \tilde{\rho}_c(k)
   ={1 \over \pi}
   +\frac{\cos{}k\ p-p^2}{\pi (p^2+1-2p \cos{}k)}
   -G_{2u}^{0}(\eta)\cos{}k+
	\cos{}k\left\{
	\begin{array}{ll}
	0& p<1 \\
	G_{2u}^{2u-2t}(\eta)+G_{2u}^{2u+2t}(\eta)& I,II\\
	a_{2t}(\eta)-a_{2t-4u}(\eta) & III
	\end{array}\right.
\end{equation}

\noindent
\underline{\sc The ferromagnetic case ($\lambda_0=0$):}
Considering the ferromagnetic case with only spin-$\uparrow$-electrons the
finite size spectrum is given by
\begin{equation}
   E=Le_\infty^\uparrow+f_\infty^\uparrow+{\pi v_c \over L}\left\{
   -{1 \over 24}+{1 \over 2}(\Delta N_c^0 -\theta_{\uparrow,p}^c)^2
	+ N_c^+ \right\}
\label{eq:fscf}
\end{equation}
and the shift $\theta_{\uparrow,p}^c$ can be given explicitely as a
function of the boundary field and the electron density (the Hubbard
interaction is not relevant in this state):
\begin{equation}
   \theta_{\uparrow,p}^c=
   -{1 \over 2}-{1 \over \pi}\arctan\left({p+1 \over p-1}\ 
     \tan{\pi n_e\over2}\right)+\theta(p-1)
\end{equation}
\section{Orthogonality exponent}
Recently, the predictions of boundary conformal field theory regarding the
relation of the finite size corrections in the spectrum of a gapless
$(1+1)$-dimensional quantum system with various boundary conditions and
scaling dimensions of certain boundary changing operators have been applied
to various problems such as Fermi edge singularities in Luttinger liquids
and the related problem of Anderson's orthogonality catastrophe in these
systems \cite{aflu:94,affl:96,fuka:96,esfr:pp}.
Here we want to apply these ideas to study the second problem, namely the
system size dependence of the overlap of the many-particle ground states
corresponding to different choices of the boundary potential.  For this we
have to consider the operator ${\cal O}_p$ switching on the boundary
chemical potential $p$.  Following Ref.~\cite{affl:96} we apply the
conformal transformation $z=L e^{\pi \omega \over L}$ to get a relation of
correlation functions in the infinite strip $\omega=u+iv$ ($0\le v\le L$
will be identified with the spatial and $u$ with the (complex) time
variable, the Fermi velocity is set to unity for this argument) with those
on the halfplane $z=\tau+ir$, $r \ge 0$.  The correlation function of the
primary boundary operator ${\cal O}_p$ in the half-plane is:
\begin{equation}
\langle AA|{\cal O}_p(\tau_1){\cal O}^\dagger_p(\tau_2)|AA\rangle
={1 \over (\tau_1-\tau_2)^{2x_p}}\ .
\end{equation}
Applying the conformal transformation we obtain the correlation function on
the strip which is given by
\begin{equation}
  \langle AA|{\cal O}_p(u_1){\cal O}^\dagger_p(u_2)|AA\rangle
  \sim  \left({\pi \over L}\right)^{2x_p} e^{-{\pi x_p \Delta u \over L}}
\end{equation}
for large $\Delta u=u_2-u_1$.  Above we denote by $|AA\rangle$ the ground
state of the system with vanishing boundary fields.  The last expression
can be evaluated by insertion of a complete set of eigenstates
$|BA;n\rangle$ of the system with chemical potential $p$ at the first site
(boundary condition `B') giving:
\begin{equation}
  \sum_n |\langle AA|{\cal O}_p|BA;n\rangle|^2 e^{-[E_n^{BA}-E_0^{AA}]\Delta u}
  \sim  \left({\pi \over L}\right)^{2x_p} e^{-{\pi x_p \Delta u \over L}}
\label{eq:cft}
\end{equation}
For the operator considered here the form factor $\langle AA|{\cal
O}_p|BA;0\rangle$ is non--zero and the exponent $x_p$ can be read off to be
\begin{equation}
   x_p={L \over \pi}\left(E_0^{BA}-E_0^{AA}\right).
\label{eq:x}
\end{equation}
From (\ref{eq:cft}) we can identify $x_p$ as the orthogonality exponent:
\begin{equation}
  |\langle AA|{\cal O}_p|BA;0\rangle|=|\langle p|0\rangle|
	\sim \left({1 \over L}\right)^{x_p},
\end{equation}
where $|p\rangle$ is the ground state of the system with boundary chemical
potential $p$.

Using the results of the previous section we can now calculate this
exponent from the finite size spectrum (the necessary generalization from
(\ref{eq:x}) to the present case of a two component Luttinger liquid with
different Fermi velocities in the respective sectors is completely
analogeous to the one in the periodic Hubbard model \cite{frko:90}).
The key to the correct identification of the orthogonality exponent is the
correct choice of $\Delta N^0_{c,s}$ in (\ref{fsopen:h}): as discussed
above the ground state energy $E_0^{AA}$ is obtained by taking $\Delta
N^0_{c,s}= \theta^{c,s}_{p=0}$.  If we compare this energy to $E_0^{AB}$ it
is crucial to compute the finite size corrections with respect to the {\em
same} reference state.  Since $|0\rangle$ and $|p\rangle$ need to be states
with the same particle numbers $N_e$ and $N_\downarrow$ this implies that
the correct choice of $\Delta N^0_{c,s}$ in $E_0^{AB}$ is again
$\theta^{c,s}_{p=0}$.

That this choice gives indeed the desired answer is checked most easily in
the ferromagnetic case: From (\ref{eq:fscf}) we obtain
\begin{equation}
  x_p={1 \over 2}\left(\theta_{\uparrow,p=0}^c-
      \theta_{\uparrow,p}^c\right)^2={1 \over 2}\left(
      {1\over\pi}\arctan\left({p+1\over p-1}\ \tan{\pi n_e\over2} \right)
      +{n_e \over 2}
      -\theta(p-1)\right)^2
\label{eq:xf}
\end{equation}
approaching $x_p={1 \over 2}(n_e-1)^2$ in the limit $p \to \infty$.  In
this ferromagnetic case the many--particle wave function is simply a slater
determinant of the one--particle functions $\Psi_k(x)\propto\sin(kx)-p
\sin(k(x-1))$. The product $\langle p|0 \rangle$ can be evaluated
numerically for finite systems leading to exponents which are in perfect
agreement with (\ref{eq:xf}).

\begin{figure}[ht]
\begin{center}
\leavevmode
\epsfxsize=0.5\textwidth
\epsfbox{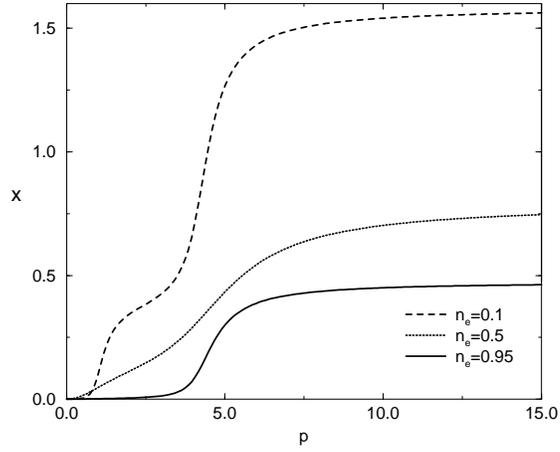}
\end{center}
\caption{Orthogonality exponent $x$ as a function of the boundary
potential $p$ for several electron densities and $u=1$.
\label{fig:expU4}}
\end{figure}

For the case of vanishing bulk magnetic field the finite size corrections
are given by (\ref{fsopen:h0}).  Choosing $\Delta N^0_{c}=\theta^{c}_{p=0}
=2\theta^{s}_{p=0} =2\Delta N^0_{s}$ we find that there are no corrections
from the spinon sector and the orthogonality exponent becomes
\begin{equation}
  x_p={1 \over 2\xi^2}\left(\theta_{p=0}^c-\theta_{p}^c\right)^2
\end{equation}
with $\theta_{p}^c$ given in (\ref{eq:tpc}). For very large $p \to \infty$
we obtain $x_p={1 \over 2\xi^2}\left(2-n_e\right)^2$.  As we approach half
filling $n_e \to 1$ the exponent becomes $x_p={\theta(p-p_2) \over 2}$.  In
Fig.~\ref{fig:expU4} we present numerical data for $x_p$ as a function of
$p$ for several values of $n_e$ and $u=1$.

\begin{figure}[ht]
\begin{center}
\leavevmode
\epsfxsize=0.5\textwidth
\epsfbox{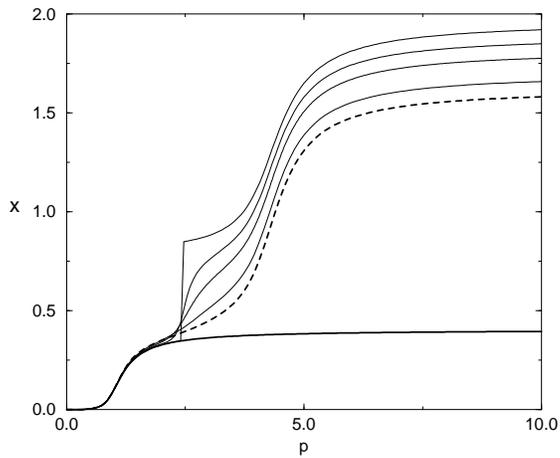}
\end{center}
\caption{
Orthogonality exponent $x$
as a function of the  boundary potential $p$ for electron
density $n_e=0.1$ and $u=1$. The bold line is the exponent for
the ferromagnetic case. The other ones have different magnetic
fields $h$, starting with $h=0$ (dashed line) up to the critical
magnetic field $h_c$ (largest exponent for $p \to \infty$)  .
\label{fig:mag}}
\end{figure}

In the general case of nonvanishing magnetic fields the exponent is given
as the sum of the respective charge and spin part $x_p=x_c+x_s$ with
\begin{eqnarray}
  x_c&=&{1\over{2\det^2(Z)}} 
                \left[\left(\theta^c_{p=0}-\theta^c_p\right)Z_{ss}
		- \left(\theta^s_{p=0}-\theta^s_p\right)Z_{cs}\right]^2 
\nonumber \\
  x_s&=&{1\over{2\det^2(Z)}} 
		\left[\left(\theta^s_{p=0}-\theta^s_p\right)Z_{cc}
		- \left(\theta^c_{p=0}-\theta^c_p\right)Z_{sc}\right]^2
\end{eqnarray}
Again, this expression simplifies for $p \to \infty$:
\begin{equation}
  \lim_{p \to \infty}x_p=\frac{\left((2-n_e)Z_{ss}-Z_{cs}\right)^2
  +\left(Z_{cc}-(2-n_e)Z_{sc}\right)^2}{{2\det^2(Z)}}.
\end{equation}
In Fig.~\ref{fig:mag} the exponent $x_p$ is shown as a function of $p$
for several magnetic fields $h$.

\begin{figure}[ht]
\begin{center}
\leavevmode
\epsfxsize=0.5\textwidth
\epsfbox{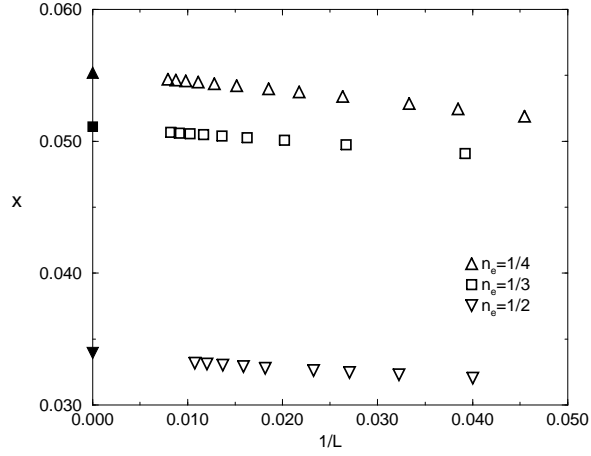}
\end{center}
\caption{ Numerical results for the exponent $x_{0.8,0.5}$ in the
ferromagneitc case for three electron densities and $u=1$ as a function of
$1/L$. The plotted exponent is calculated from the ratio $\langle
p_1p_L|00\rangle|_L / \langle p_1p_L|00\rangle|_{L+1/n_e}$.
\label{fig:num}}
\end{figure}

\begin{figure}[ht]
\begin{center}
\leavevmode
\epsfxsize=0.5\textwidth
\epsfbox{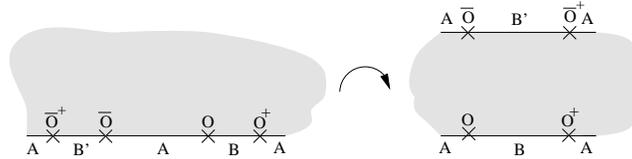}
\end{center}
\caption{
Conformal mapping of the infinite half plane with 
$\omega={ L \over \pi}\ln({z \over L})$ to the strip.
\label{fig:conf}}
\end{figure}

Finally, let us remark on the effect of a second boundary potential $p_L$
at site $L$: the BAE (\ref{eq:bae}) are modified by another factor
$s_{p_L}(k)$ leading to additional shifts $\theta^{c,s}_{p_L}
-\theta^{c,s}_{p_L=0}$ in the expressions for the finite size spectrum
(\ref{fsopen:h}).  In this case the orthogonality exponent $x_{p_1p_L}$
\begin{equation}
\langle p_1p_L|00\rangle \sim \left({1 \over L}\right)^{x_{p_1p_L}},
\end{equation}
can not be obtained by simply adding the new shifts.  Instead, numerical
studies of the ferromagnetic case (see Fig.~\ref{fig:num}) suggest that the
exponent $x_{p_1p_L}$ is given by
\begin{equation}
   x_{p_1p_L}=x_{p_1}+x_{p_L}
\label{eq:sum}
\end{equation}
i.e.\ the effects from the two boundaries are additive.  In the framework
of boundary conformal field theory this result is a consequence of the fact
that changing the potential at {\em both} boundaries is not possible by the
action of a single boundary changing operator ${\cal O}_{p_1p_L}$ but
rather two operators ${\cal O}_{p_1}$ and $\overline{\cal O}_{p_L}$ as
becomes obvious when one switches back from the system on the strip to that
on the half-plane (see Fig.~\ref{fig:conf}).  Hence, the correlation
function considered is
\begin{equation}
  |\langle AA|{\cal O}_{p_1}(\tau'_1)
    \overline{{\cal O}}_{p_L}(\tau_1)
    \overline{{\cal O}}^\dagger_{p_L}(\tau_2)
    {\cal O}^\dagger_{p_1}(\tau'_2)|AA\rangle|
\end{equation}
which gives (provided that $|\tau_i-\tau'_i| \ll |\tau_1-\tau_2|$)
\begin{equation}
   |\langle AA|{\cal O}_{p_1}(\tau'_1)
	{\cal O}^\dagger_{p_1}(\tau'_2)|AA\rangle| 
   |\langle AA|\overline{{\cal O}}_{p_L}(\tau_1) 
	\overline{{\cal O}}^\dagger_{p_L}(\tau_2)|AA\rangle|
   = \frac{1}{(\tau'_1-\tau'_2)^{2x_{p_1}}}\
     \frac{1}{(\tau_1-\tau_2)^{2x_{p_L}}}
\end{equation}
for the leading asymptotic of the correlator in the semiinfinite plane.
Conformal mapping of this expression to the strip results in (\ref{eq:sum}).

\section*{Acknowledgements}
This work has been supported by the Deutsche Forschungsgemeinschaft under
Grant No.\ Fr~737/2--2.


\setlength{\baselineskip}{13pt}

\newpage
\section*{Figures}

\end{document}